# Surface data imputation with stochastic processes

A Jawaid*, S Schmidt, M Lotz and J Seewig

Institute for Measurement and Sensor Technology, University of Kaiserslautern-Landau, Kaiserslautern, Germany

E-mail: arsalan.jawaid@rptu.de

## Abstract

Spurious measurements frequently occur in surface data from technical components. Excluding or ignoring these spurious points may lead to incorrect surface characterization if these points inherit features of the surface. Therefore, data imputation must be applied to ensure that the estimated data points at spurious measurements do not deviate strongly from the true surface and its characteristics. Traditional surface data imputation methods rely on simple assumptions and ignore existing knowledge of the surface, resulting in suboptimal estimates. In this paper, we propose the use of stochastic processes for data imputation. This approach, which originates from surface texture simulation, allows a straightforward integration of *a priori* knowledge. We employ Gaussian processes with both stationary and non-stationary covariance structures to address missing values in surface data. In addition, we apply the method to a real-world scenario in which a spurious turned profile is obtained from an actual measurement. Our results demonstrate that the proposed method fills the missing values by maintaining the surface characteristics, particularly when surface features are missing.

## 1. Introduction

A valid characterization of technical surfaces requires reliable and accurate measurement data. However, surface measurement data always face heteroskedastic uncertainties that depend on the measurement principle and the surface itself [1, 2]. Such uncertainties can lead to false or spurious data if measurements significantly deviate from the ground truth, thereby impairing the evaluation process and causing incorrect characterization of surface features [2, 3]. Additionally, post-processing of raw measurement data can introduce spurious surface data. For example, white-light interferometry may produce artifacts such as batwings at steps [4] or ghost steps that do not exist on the surface due to erroneous data processing [5].

Spurious data must be detected and masked during the evaluation process if present in the measurement [6]. This prevents the mischaracterization of surfaces, but only if surface features are not mistakenly ignored. Masking out actual features can lead to incorrect surface characterization [2]. Therefore, models should ideally estimate values at spurious locations considering observed data, ensuring that the estimated points closely resemble the actual surface. We distinguish two scenarios in surface measurements:

(i) Estimating intermediate values,
(ii) filling missing values [7].

The first problem is known as data interpolation, while the second is hereinafter referred to as data imputation. In surface topography literature, these terms are often used interchangeably, although there is an awareness that they

* Author to whom any correspondence should be addressed.

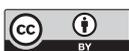

represent distinct problems. For post-measurement analyzes, for instance, it is advantageous to map a surface measurement to an evenly spaced grid [3]. In these contexts, we assume that no features are missing and that the data is adequately distributed over the measurement area of interest. Thus, data interpolation typically addresses irregularly spaced data in surface measurements. Surface data imputation, although similar, can be more complex. This problem arises when spurious points coincide with surface features, necessitating more expressive models, whereas data interpolation can be performed with basic models.

In the surface topography literature, the imputation of missing data at spurious locations is generally handled in two ways [8]. The missing values are filled with a reference value or inferred based on valid neighboring data [8].

The first approach is referred to as mean or median imputation if the reference value is the mean or median of the valid data. Although this technique is simple and efficient, it assumes a planar surface at spurious locations. Since missing values in surface data rarely align with planar structures, this method is generally not preferred [8].

The second approach, local regression-based imputation, estimates missing values by using adjacent valid data in a regression model. For example, a smooth regression model (not explicitly stated) has been applied in [2]. A common example is nearest-neighbor mean imputation, where a missing point is assigned the mean value of its valid neighbors [9]. This method is effective for isolated missing points, but fails when entire regions are missing. This creates plateaus that may not reflect the actual surface.

To improve the imputation of missing regions, a median filter can be applied [6, 10]. This method fills in missing regions by applying the filter repeatedly [6, 10]. In practice, any filter can be used for surface data imputation. They are effective at preserving edges [6, 10], but often smooth out small-scale details. Another approach is weighted interpolation [8, 11]. Here, missing values are filled with a regression-based model, where adjacent valid points are weighted based on their distance to the spurious location [8, 11]. Even though both approaches, filtering and weighted interpolation, consider spatial relationships better than simpler approaches, they rely on local information. As a result, they often fail to capture broader spatial dependencies.

To address these limitations, stochastic models such as ordinary kriging [12] have gained attention for surface data imputation. They estimated missing values by modeling spatial relationships within the surface data [12]. Their methods offer improvements over deterministic approaches; however, they assume simple stationary covariance structures. This limits their ability to represent complex surface structures. Other approaches [13–15] also use simple stationary covariance structures, but improve accuracy with additional data and model resources.

We propose using Gaussian processes (GPs) with both stationary and non-stationary covariance structures to address surface data imputation. Our method falls into the category of stochastic models, but differs in its ability to explicitly incorporate *a priori* knowledge about the surface into the model. Furthermore, it does not require extra data or model resources, making it a practical and efficient solution.

We apply this approach to artificially generated profile traces, where missing features are introduced through watershed segmentation [16] or due to optical limitations of a virtual measurement [17]. In addition, we evaluate our method on a real measurement, where spurious data are introduced by specifically configuring the measurement process.

In this paper, we demonstrate that our approach enables the reconstruction of missing surface features without requiring additional data. Compared to standard surface data imputation methods, our approach preserves fine-scale surface details and effectively captures non-stationary variations.

## 2. Gaussian process model

### 2.1. Problem setting

This paper focuses only on surface data imputation. Therefore, we differentiate its problem setting from that of data interpolation. Our notation follows a structure similar to that in [18].

An ideal surface measurement has $N$ data points $z^* \in \mathcal{Z} \subset \mathbb{R}^N$ on an evenly spaced grid $X^* \in \mathcal{X} \subset \mathbb{R}^{N \times D}$. We denote a profile trace of a surface with $D = 1$ and, for simplicity, describe the problem setting only for profiles. It is worth noting that extending this to areal surfaces ($D = 2$) is straightforward, but data imputation for areal surfaces is beyond the scope of this paper. For $D = 1$, the grid is defined as

$$X^* = \{x_i^*\}_{i=1}^N, \quad x_i^* = x_0 + i\Delta\xi, \tag{1}$$

where $\Delta\xi > 0$ is the sampling spacing and $x_0$ is typically set to zero. In addition, we denote a real surface measurement with $M$ consecutive data points as $S = \{(\tilde{x}_j, \tilde{z}_j)\}_{j=1}^M$.

*2.1.1. Data interpolation.* We assume that no features are missing and that the surface measurement $S$ is sufficiently distributed over the sample space $\mathcal{X}$. The goal is to find a function that returns the intermediate values at the grid locations $\tilde{X} = X^* \setminus \{\tilde{x}_j \mid j = 1, \ldots, M\}$

$$f(\tilde{x} \mid S) = \hat{z}, \quad \tilde{x} = (x)_{x \in \tilde{X}}, \tag{2}$$

such that $\hat{z} \approx z^*$ holds. The grid locations that are not present in the surface measurement are denoted in the vector $\tilde{x}$.

*2.1.2. Data imputation.* The measurement data $S$ have spurious locations at $X_m \subset \{\tilde{x}_j\}_{j=1}^M$. In surface data imputation, we do not assume that no features are missing. The goal is to find a function that fills in the missing measurements

$$f(\boldsymbol{x}_m \mid S_a) = \hat{\boldsymbol{z}}, \quad \boldsymbol{x}_m = (x)_{x \in X_m}, \tag{3}$$

where $\boldsymbol{x}_m$ is a vector containing all missing locations in the data, $S_a$ represents the valid surface measurements at locations $X_\mathrm{a} = \{\tilde{x}_j\}_{j=1}^{M} \setminus X_\mathrm{m}$, and $\hat{z}$ resembles ideally the non-spurious surface values. In the following, we represent $z_\mathrm{a}$ as the vector of valid surface heights at valid locations, denoted as vector $\boldsymbol{x}_\mathrm{a} = (x)_{x \in X_\mathrm{a}}$.

*2.2. Model*

A stochastic process that has finite-dimensional Gaussian distributions is referred to as a GP [19]. Similarly to a Gaussian distribution, a GP $G(x)$ is characterized by an expectation function $\mu(\cdot)$ and an autocovariance function $r(\cdot,\cdot)$, expressed as $G(x) \sim \mathcal{GP}(\mu, r)$. The expectation and autocovariance functions are defined as follows

$$\mu(x) = \mathbb{E}[G(x)],$$
$$r(x,x') = \mathbb{E}\left[(G(x) - \mu(x)) \cdot \overline{(G(x') - \mu(x'))}\right]. \quad (4)$$

According to [20], a GP combined with a noise model

$$\hat{\boldsymbol{Z}} \mid \boldsymbol{g} \sim p(\hat{z} \mid \boldsymbol{g}), \quad (5)$$

can be used to model rough surfaces. This noise model may account for measurement noise or surface defects resulting from non-stationary manufacturing processes, such as chip formation. The expectation function, the autocovariance function, and the noise model might have model parameters $\boldsymbol{\alpha}$ [19, 20].

In the literature, GPs have been used to model and simulate surface textures [20–22]. These models can replicate surface characteristics by appropriately incorporating prior knowledge into the expectation and autocovariance functions [19, 20]. Consequently, we also adopt a GP approach for surface data imputation. We model the function (3) by transparently embedding prior knowledge into the GP functions and filling missing values by inference. This multivariate (posterior) Gaussian distribution is conditioned on valid measurement data $S_\mathrm{a}$. For an additive Gaussian noise model, the function is defined as follows

$$f(\boldsymbol{x}_\mathrm{m} \mid S_\mathrm{a}) = \mathcal{N}\left(\hat{z} \mid \hat{\boldsymbol{\mu}}, \hat{\boldsymbol{\Sigma}}\right),$$
$$\hat{\boldsymbol{\mu}} = \mu(\boldsymbol{x}_\mathrm{m}) + \boldsymbol{\Sigma}_{\mathrm{m,a}} (\boldsymbol{\Sigma}_{\mathrm{a,a}} + \boldsymbol{\Omega})^{-1} (z_\mathrm{a} - \mu(\boldsymbol{x}_\mathrm{a})), \quad (6)$$
$$\hat{\boldsymbol{\Sigma}} = \boldsymbol{\Sigma}_{\mathrm{m,m}} - \boldsymbol{\Sigma}_{\mathrm{m,a}} (\boldsymbol{\Sigma}_{\mathrm{a,a}} + \boldsymbol{\Omega})^{-1} \boldsymbol{\Sigma}_{\mathrm{a,m}},$$

where $\boldsymbol{\Sigma}_{\cdot,\cdot}$ are matrices element-wise constructed by the prior autocovariance function (4), and $\boldsymbol{\Omega}$ is a covariance matrix derived from the noise model. For example, if the noise model is additive white Gaussian noise, the noise matrix is $\boldsymbol{\Omega} = \sigma_\mathrm{n}^2 \boldsymbol{I}$. In addition to this case, we can also introduce additive colored Gaussian noise

$$[\boldsymbol{\Omega}]_{ij} = \sigma_\mathrm{n}^2 \exp\left(-\frac{1}{2} \frac{(x_i - x_j)^2}{\theta_\mathrm{n}^2}\right). \quad (7)$$

Roughness profiles are treated as samples of the Gaussian distribution [20, 22] rather than the posterior expectation mean.

Therefore, we applied surface data imputation by posterior sampling of the topography heights at those spurious locations.

Using this approach, we address data imputation for two simulated use cases: a turned profile and a chirp-structured profile. In addition, we apply the method to a real-world scenario in which a spurious turned profile is obtained from an actual measurement. A detailed description of the data acquisition process for each use case is provided in the results section. For now, we assume that the profiles are given and contain missing data.

The autocovariance function plays a crucial role in the model, as it encodes key surface characteristics such as smoothness and periodicity [19, 20, 23]. In contrast, the expectation function mainly accounts for deterministic trends in the surface [19]. However, assuming a constant expectation function (e.g. zero) does not significantly constrain the model, as the posterior mean can still adapt to the data [19]. Based on this discussion, our approach prioritizes the autocovariance function to characterize surfaces. The specific autocovariance models used for imputation are detailed in the following section. In all use cases, the expectation function is set to $\mu(\cdot) = 0$, and the noise model is assumed to be additive white Gaussian noise.

*2.2.1. Turned profile.* Turned profiles can be reasonably well represented with stationary autocovariance functions (see [20, 24]). Therefore, we modeled the autocovariance function using the stationary spectral mixture model since it has been successful in modeling periodic patterns, has a simple form, and is interpretable [23]. It may be tempting to use more expressive autocovariance functions for turned profiles, such as the non-stationary spectral mixture model [25]. However, this generalization comes with a complex form that makes the parameter space vast and optimization more challenging. Thus, is the stationary assumption sufficiently appropriate, as is for turned profiles, the stationary spectral mixture model is adequate. The autocovariance function is given by [23]

$$r_\mathrm{SM}(\tau) = \sum_{k=1}^{Q} w_k \cos(2\pi \tau f_k) \exp(-2\pi^2 \tau^2 v_k), \quad (8)$$

where $\tau = x - x'$ holds, and $\{(w_k, f_k, v_k)\}_{k=1}^{Q}$ are the parameters of the autocovariance function. This autocovariance function corresponds to Gaussian mixtures in the Fourier domain [23]. Thus, $w_k$ is the weight, $f_k$ is the mean, and $v_k$ is the variance of the $k$th Gaussian [23].

Although the model is general, prior knowledge of turned profiles or other structured profiles can be incorporated relatively easily. The parameters of the autocovariance function can be interpreted as follows. The parameter $f_k$ specifies the expected profile element frequency. If the surface has many profile elements, $w_k$ factors in the contribution of each profile element and all factors $w_k$ sum up to the profile height variance. Lastly, $v_k$ can be interpreted as a variation of the pro-file element frequency. In limiting cases, $v_k \to 0$ implies an ideal profile spacing, while $v_k \to \infty$ blurs the profile element spacing so that it can have any value in $(0, \infty)$ with uniform probability.

For spurious turned profiles, initial estimates of the roughness parameters, the mean profile element spacing $R_{sm}$ and the root mean square height $R_q$, can be used to include prior knowledge. For our case, we initialized two model parameters by setting $f_k = 1/R_{sm}$ and $w_k = R_q^2$ for one $k$ of the Gaussian mixture, respectively. For surface structures with complex periodic patterns, superimposed profile elements can be handled by increasing the number of mixtures in the model. In such cases, it is important to incorporate prior knowledge individually to ensure accurate representation and optimization.

The model parameters, including the noise parameter, are $\boldsymbol{\alpha} = \{\sigma_n^2\} \cup \{(w_k, f_k, v_k)\}_{k=1}^Q$, where we set the number of Gaussian mixtures $Q$ to 5, as in [20]. These parameters are optimized for the respective data set by maximizing the marginal log likelihood [19]

$$J(\boldsymbol{\alpha}) = \log p(z_a \mid \boldsymbol{\alpha}). \tag{9}$$

*2.2.2. Chirp-structured profile.* To demonstrate the capability of our approach to solve more complex problems by making use of prior knowledge, we address surface data imputation in the case of a chirp-structured profile. This profile has varying wavelengths and is inherently non-stationary, necessitating the use of a non-stationary model. The generalized spectral mixture model is a non-stationary extension of the spectral mixture model [25]. Similarly to superimposed stationary surface structures, superimposed non-stationary structures can be handled by increasing the number of mixtures in the model. Given its successful application in extrapolating a chirp-like signal with a single mixture component in [25], we also use only one mixture component for our problem, that is

$$r_{GSM}(x, x') = w(x) w(x') r_G(x, x') \\ \cdot \cos(2\pi(f(x)x - f(x')x')), \tag{10}$$

with the Gibbs autocovariance function [26]

$$r_G(x, x') = \sqrt{\frac{2\lambda(x)\lambda(x')}{\lambda(x)^2 + \lambda(x')^2}} \\ \cdot \exp\left(-\frac{(x-x')^2}{\lambda(x)^2 + \lambda(x')^2}\right), \tag{11}$$

where $\log w(x) \sim \mathcal{GP}(\mu_C, r_{SE})$, $\log \lambda(x) \sim \mathcal{GP}(\mu_C, r_{SE})$, and logit $f(x) \sim \mathcal{GP}(\mu_C, r_{SE})$ are independent GPs with constant expectations, referred to as latent GPs. The log transformation ensures that the latent functions remain positive, while the logit transformation restricts the frequencies to positive values and keeps them below the Nyquist frequency, as described

in [25]. Each latent GP has its own squared exponential autocovariance function

$$r_{SE}(x, x') = \sigma_k^2 \exp\left(-\frac{1}{2} \frac{(x - x')^2}{\theta_k^2}\right). \tag{12}$$

Furthermore, latent GPs have representatives at arbitrarily selected locations $\boldsymbol{x}_L = (x_i)_{i=1}^P$. For example, $\bar{\boldsymbol{w}} = (\log w_i)_{i=1}^P$ are additional parameters of the model. To make inferences on the overall model, we first approximate the latent representatives at the required locations (e.g. $\log w(\boldsymbol{x}_m)$) with the standard Gaussian posterior mean [27, 28]. Secondly, we include the approximated latent representatives in (10) to make finally the overall inference (6). This model allows us to include information on wavelengths in the latent frequency model $f(x)$ using the representatives or the latent GP functions itself.

We set the locations of all latent representatives on an evenly spaced grid with $P = 100$ points that cover the measurement area. The parameters of the model are $\boldsymbol{\alpha} = \{\sigma_n^2\} \cup \{\bar{\boldsymbol{w}}, \bar{\boldsymbol{\lambda}}, \bar{\boldsymbol{f}}\} \cup \{\sigma_{k,h}^2, \theta_{k,h}\}_{h \in \{w, \lambda, f\}}$. The model parameters are selected by maximizing the marginal log posterior, as in [25, 27, 28]

$$J(\boldsymbol{\alpha}) \propto \log p(z_a \mid \boldsymbol{\alpha}) + \log p(\boldsymbol{\alpha}). \tag{13}$$

We also applied whitening to latent representatives to speed-up gradient-based optimizing [25, 28, 29]. Unlike previous works, we adaptively whiten the parameters since the latent model parameters change during optimization.

## 3. Results

We address data imputation problems for profile traces, using both simulated and real measurement data. In the simulated cases, spurious data are artificially introduced. In the real case, the raw measurement data are configured to produce many spurious data naturally. Each use case is structured as follows. First, the data acquisition process is described. Next, the surface data imputation method (see section 2.2) is applied. Finally, the results are discussed.

Inference and optimization are computationally expensive and scale with the number of valid surface data [19]. For the simulated turned profile, we approximate the autocovariance function (8) using structured kernel interpolation [30]. After model selection, posterior sampling for data imputation is performed using the Lanczos variance estimate [31]. In contrast, for the simulated chirp profile and the real turned profile, the number of valid data points is relatively small. As a result, approximations are unnecessary, and sampling is done with the Cholesky decomposition. A summary of the data sizes is provided in table 1.

We performed our experiments using GPyTorch [32]. All computations were executed on an Intel® Core™ i7-10 750 H CPU with 32 GB (2 × 16 GB) of DDR4 3200 MHz RAM.

**Table 1.** Summary of data sizes for the data imputation problem. Each use case is presented with its total size (simulation size if considered), the number of spurious locations, and the size of valid data points. The chirp data refers to the profile trace used in the chirp simulation, not the areal surface.

|  | Total | Spurious | Valid |
|---|---|---|---|
| Simulated turned profile | 8000 | 953 | 7047 |
| Simulated chirp-structured profile | 5000 | 416 | 801 |
| Real turned profile | 1592 | 622 | 970 |

### 3.1. Simulated turned profile

This section addresses the data imputation problem using a simulated turned profile. The simulated profile is generated with the GP approach [20, 24]. Spurious measurements are emulated using the profile-based watershed segmentation method [16].

#### 3.1.1. Data acquisition.
We followed the method in [24] to simulate a turned profile. They used a periodic autocovariance function [24, 33]

$$r(x,x') = \sigma_{\text{sim}}^2 \exp\left(-\frac{1}{2}\frac{\sin^2\left(\frac{\pi}{\lambda_{\text{sim}}}(x-x')\right)}{\theta_{\text{sim}}^2}\right), \quad (14)$$

where $\lambda_{\text{sim}}$ specifies the period, $\sigma_{\text{sim}}^2$ represents the variance, and $\theta_{\text{sim}}^2$ controls the profile variation. The expectation function is set to zero, and the noise model is the additive colored Gaussian noise (7). For further details, see [24]. A section of the simulated turned profile, along with the corresponding simulation parameters, is shown in figure 1. The parameters were chosen to imitate a real turned profile.

The simulation of the turned profile does not yet contain spurious data. According to [34], spurious data often appear at high surface gradients in white-light interferometer measurements. To emulate this effect, we introduce spurious data into the simulated profile at locations with high gradients. To formulate the data imputation problem, we label entire dales of the profile with the smallest (local) widths as spurious. This approach assumes that, for dales of equal depth, smaller widths correspond to steeper gradients around the pit. Although spurious data typically occur at high gradients rather than specifically near the pit, we remove entire dales to increase the complexity of the imputation task. In particular, we segmented the profile into features (dales) using profile-based watershed segmentation [16]. Our segmentation process performed pruning based on the local volume of the dales and selected those with the smallest local width. The pruning threshold is selected to achieve optimal periodicity, ensuring that the segments are as equal in size as possible. For more information, refer to [16].

#### 3.1.2. Data imputation.
Prior knowledge was incorporated by initializing the model parameters $f_k$ and $w_k$ of a $k$th Gaussian using the profile element spacing and the root mean square height, as described in section 2.2. Since the turned profile was simulated with the expected profile element spacing and root mean square height, we set $f_k = 1/\lambda_{\text{sim}}$ and $w_k = \sigma_{\text{sim}}^2 + \sigma_{\text{n}}^2$.

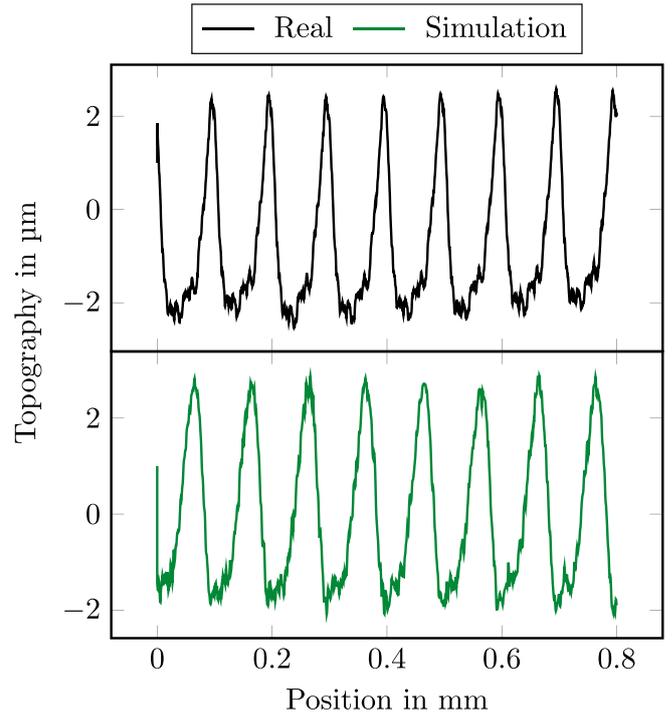

**Figure 1.** A real turned profile and a section of the simulated turned profile is shown. The real profile was measured with a tactile measurement device. The simulated profile was generated with model parameters: $\sigma_{\text{sim}}^2 = 10\,\mu\text{m}^2, \theta_{\text{sim}} = 0.8, \lambda_{\text{sim}} = 0.1\,\text{mm}, \sigma_{\text{n,sim}}^2 = 0.02\,\mu\text{m}^2, \theta_{\text{n,sim}} = 0.001\,\text{mm}$, and $N = 8000$ sampling points with a sampling distance $\Delta x = 5 \times 10^{-4}\,\text{mm}$.

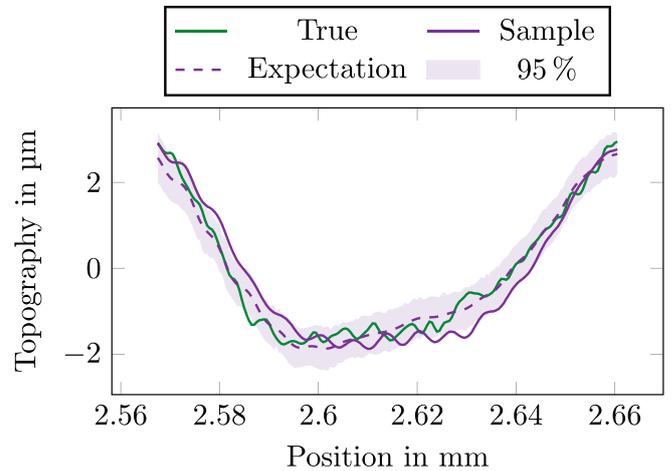

**Figure 2.** Estimated profile feature obtained from a sample of the posterior GP. The posterior GP is represented by the posterior expectation and the 95% uncertainty range.

Eventually, we reconstructed the missing values after optimizing the autocovariance function and the noise model parameters. Figure 2 shows an example of an estimated feature using the posterior GP. The imputation demonstrates that missing features can be recovered, with almost all true surface values falling within the 95% uncertainty range of the posterior. For better visualization, figure 3 presents the complete spurious profile after data imputation using a posterior sample of the GP. The spurious profile features are successfully reconstructed, as the stationary spectral mixture model was able to approximate the original simulation model (14). This was possible because prior information was available and sufficient valid dales provided additional contextual data for the reconstruction. However, this approach can introduce jumps or discontinuities in which spurious regions meet valid data points. For instance, around 0.5 mm in figure 3, a jump appears on the right side of the spurious region as the imputed data (sample) transition to the valid data. This discontinuity arises because we use an additive white Gaussian noise model, which introduces a constant variance term $\sigma_n^2$ at all positions. Consequently, the posterior variance never fully vanishes, even near known data points. Figure 2 shows this behavior: the uncertainty range remains near valid data (e.g. at $\approx 2.66$ mm). Technically, the topography heights at these locations have a non-zero posterior variance. As a result, samples drawn from the posterior follow this distribution and yield to visible jumps. To address this issue, a heteroscedastic Gaussian noise model could be used instead of white Gaussian noise. This approach allows the noise variance to adapt based on data availability, as proposed in [27, 35, 36]. Note that this approach formally models the topography height $Z(x)$ as a non-stationary model.

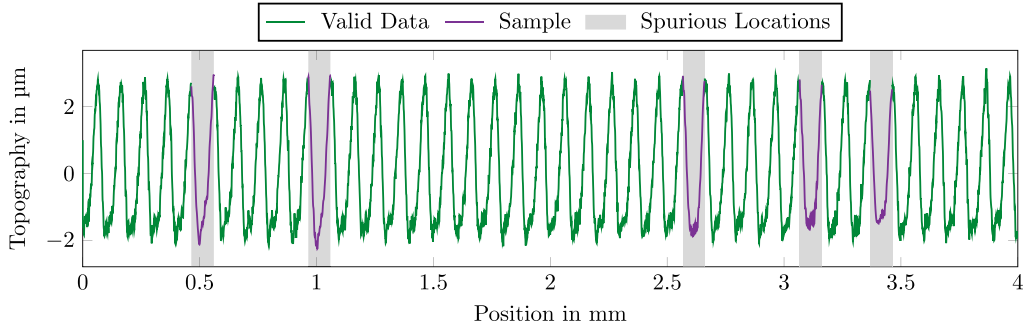

**Figure 3.** A synthetically generated spurious measurement based on simulation data. Spurious locations are highlighted by shaded areas in the dales with the five lowest local widths. The dales and local widths are segmented using watershed segmentation with pre-pruning by volume [16]. Surface data imputation is performed using a sample from the posterior GP.

### 3.2. Simulated chirp-structured profile

In this second use case, we simulate a chirp-structured surface and perform a virtual measurement to introduce spurious data. Finally, we extract a profile trace from the generated areal spurious measurement and apply surface data imputation.

*3.2.1. Data acquisition.* The chirp-structured surface is a cosine wave with increasing wavelength after each period. We slightly modified the parameters of this chirp-structured surface compared to those used in other chirp material measures [37–39]. This modification allows us to achieve high gradients, which increases the likelihood of producing spurious data during virtual measurement. The chirp profile is defined as

$$\mu_{\text{Chirp}}(x) = \frac{a}{2}\cos(2\pi x/\lambda(x)),$$

$$\lambda(x) = \boldsymbol{\lambda}_k \quad \text{for} \quad \sum_{m=1}^{k}\boldsymbol{\lambda}_{m-1} \leqslant x < \sum_{m=0}^{k}\boldsymbol{\lambda}_m, \qquad (15)$$

where the (double) amplitude is set to $a = 5\,\mu\text{m}$, and the wavelengths are $\boldsymbol{\lambda} = (10^{1+k/k_{\max}}\,\text{mm})_{k=0}^{k_{\max}}$ with $k_{\max} = 24$ [37].

Yet, the chirp surface we have looked at so far is almost surely smooth. To make the surface more realistic, we introduced small correlated defects by adding colored noise (7) to the latent chirp structure $\mu_{\text{Chirp}}(x)$. Additionally, measurement noise is omitted here. It should be noted that we simulated an areal chirp surface and applied data imputation on an extracted profile trace, since the virtual measurement device requires areal surface data. The areal simulation was obtained by stacking the latent chirp structure $\mu_{\text{Chirp}}(x)$ multiple times and using a rational-invariant colored noise model, where $x_i$ - $x_j$ is replaced by the Euclidean norm in (7). The resulting areal chirp-structured surface is shown in the top row of figure 4.

To emulate spurious data on the simulated chirp surface, we used a virtual measurement device that simulates an optical microscope–a focus variation instrument–via ray tracing with NVIDIA's OptiX engine [17]. The virtual model illustrates the optical limitations and emulates the corresponding spurious points. For example, when virtually measuring the chirp surface at high gradients, holes may appear in the measurement data because the sensors do not detect signals. More details on the virtual measurement device can be found in [17].

To construct the areal surface measurement, we processed the image stack returned by the virtual measurement device using MountainsMap® Premium version 7.4 software. We used the '3D reconstruction from multifocus images' function and configured the parameters to classify many measurement points as spurious. The reconstructed virtual measurement of the chirp-structured surface is shown in the bottom row of the figure 4. Since surface data imputation was performed on a profile trace, we extracted a profile from the virtual measurement (see figure 5). This figure also illustrates the deviations of the virtual measurement from the ground truth (simulated) profile trace. The virtual measurement shows stretched features, with higher hills and lower dales. We note that the virtual measurement has fewer spurious points in the horizontal midsection of the areal surface, whereas around small wavelengths the spurious points are dense. This occurs because high gradients, correlated with small wavelengths, redirect fewer light rays to the sensor. Moreover, the outer sections of the surface also have more spurious points due to the limited light rays reaching the sensor.

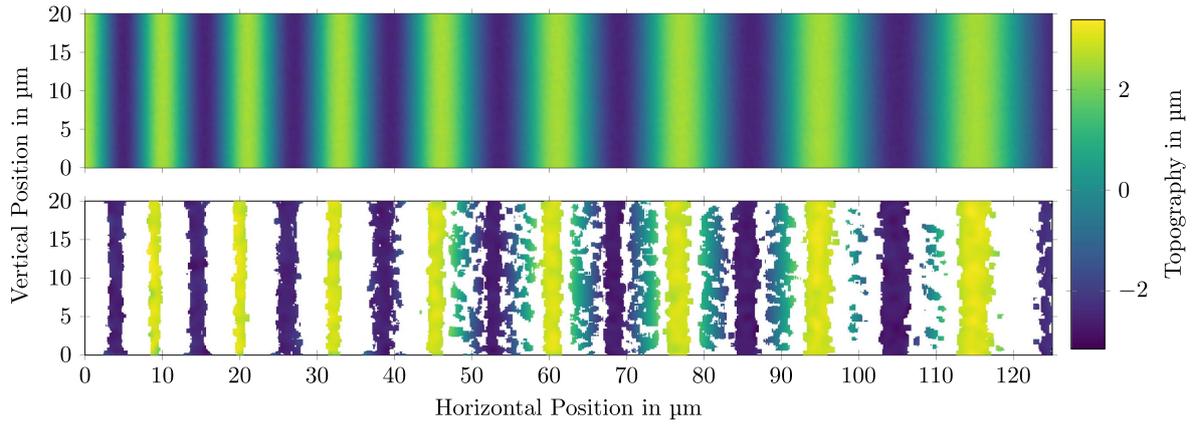

**Figure 4.** The top row shows the simulated chirp-structured areal surface with parameters $\sigma_{n,sim} = 1 \times 10^{-4}\,\mu m^2$, $\boldsymbol{\theta}_{n,sim} = (0.02\,mm, 0.02\,mm)$, and $N = 5000 \times 80$ sampling points with a sampling distance $\Delta x = 0.025\,\mu m$. The bottom row presents the virtually measured areal surface, reconstructed using MountainsMap® Premium version 7.4. We set the window size to 5 points, smoothness to 15 points, and number of points declared as spurious to many.

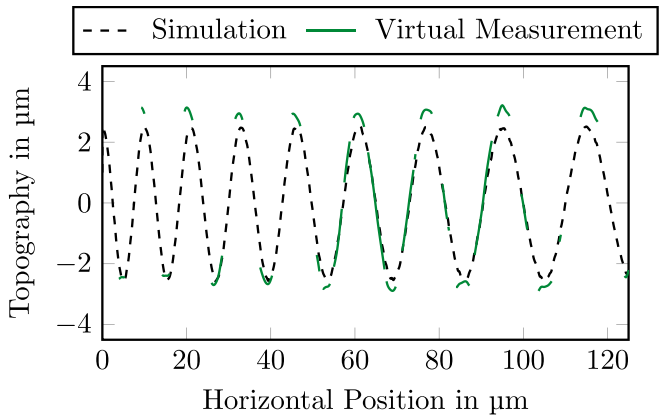

**Figure 5.** Extracted profile traces from areal chirp-structured surfaces (see figure 4) at a vertical position of 10 µm.

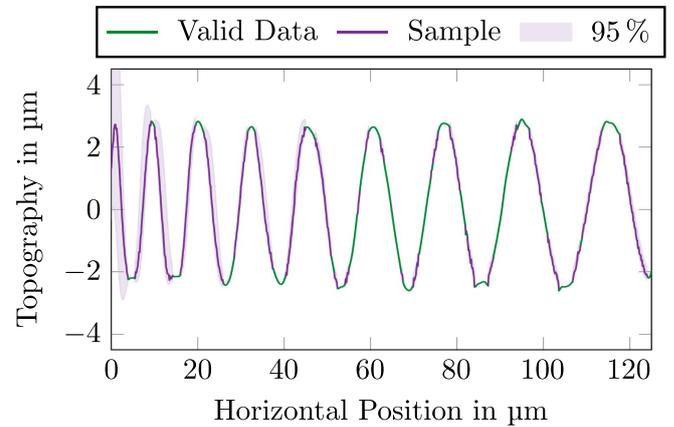

**Figure 6.** Virtually measured chirp-structured profile, reconstructed using a sample from the posterior GP. The posterior GP is given by the symmetric 95% confidence bounds.

*3.2.2. Data imputation.* We addressed the data imputation problem with the proposed model (10). The prior knowledge of increasing wavelengths was incorporated by initializing the latent frequency representatives $\bar{f}$ accordingly.

After optimizing the model parameters, the posterior GP is presented in figure 6. For clarity, the figure shows only a posterior sample and the 95 % confidence bounds of the posterior GP. The posterior GP is able to infer the varying wavelengths, particularly in regions with small wavelengths, where no data are available. As a result, the reconstructed profile closely resembles the true profile shown in figure 5. However, the true profile does not always fall within the confidence bounds, especially near the peaks and pits. This discrepancy arises because the relation between the virtual measurement and the true profile is not explicitly modeled. Extending this approach with a model that accounts for the transfer characteristics of the virtual measurement device could reduce this error. Additionally, as observed in section 3.1, jumps appear where imputed data transition to valid data. These false jumps occur because the model uses additive white Gaussian noise. Using an additive heteroscedastic Gaussian noise model may improve the imputation.

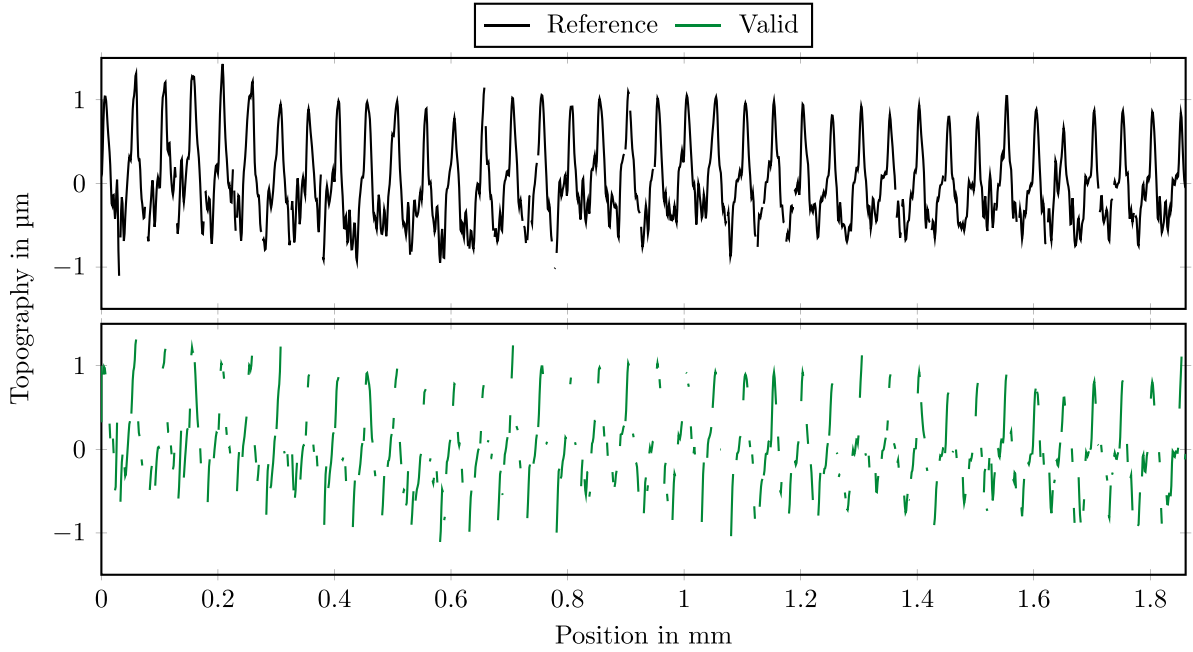

**Figure 7.** Extracted profile trace of a measured turned surface. The measurement used a white-light interferometer with an 8 bit camera and different post-processing parameters to control spurious points. The first post-processing step removed pixels from both profiles that had an intensity variation of less than 6 in the correlogram. The second step was applied only to the spurious measurement. It removed an additional 75% of remaining pixels with low envelope amplitude in the correlogram. Afterward, the noise was removed from both profiles using a linear Gaussian filter with a nesting index of $n_s = 8\,\mu m$.

### 3.3. Real turned profile

Unlike the previous sections, this part studies the data imputation problem in a real-world scenario. We analyze a real turned profile obtained through an actual measurement process, where the setup is intentionally configured to produce spurious points.

*3.3.1. Data acquisition.* The turned shaft used in this study was machined with a feed rate of 0.05 mm per revolution. The areal measurement of the surface was performed using a TopMap Micro.View®+ white-light interferometer with an 8 bit camera. To introduce spurious measurements, we selected an optical lens with 5 × magnification and a numerical aperture of 0.13. In addition, the exposure time was minimized to 1/50 000 s.

Not all pixels in the raw measurement data were considered in the final surface measurement. To determine valid pixels, two threshold parameters were subsequently applied using the Topography Measurement System Software (version 4.2). The first threshold parameter defines the minimum intensity variation in the interferometric correlogram. If the intensity variation is below this limit, the pixel is labeled as spurious. The second threshold parameter defines the minimum amplitude of the envelope of the interferometric correlogram.

For comparison, we generated a reference measurement with fewer spurious points and a spurious measurement using the same raw measurement data. In both cases, the first threshold parameter was set to an appropriate value of 6. However, to produce more spurious points in the spurious measurement, the second threshold parameter was set at a relative value of 75%. Consequently, approximately 75% of the previously valid points were declared spurious, based on the lowest envelope amplitudes. Note that the second post-processing step was not used for the reference measurement.

Each resulting areal measurement consisted of 1200 × 1592 measurement points, from which a profile trace of 1592 data points was extracted. The means of both profile traces are removed, and the noise was filtered using a linear Gaussian filter with a nesting index of $n_s = 8$ mm [40]. The spurious profile and the reference profile, which contain a relatively few spurious points, are shown in figure 7.

*3.3.2. Data imputation.* In contrast to section 3.1, the profile element spacing and root mean square height are not explicitly known in this case. Instead, we estimated the profile element spacing using the feed rate and the root mean square height using the sample standard deviation of the spurious profile. These estimates were then used to initialize the model parameters $f_k$ and $w_k$ of a *k*th Gaussian in the turned profile model (see section 2.2).

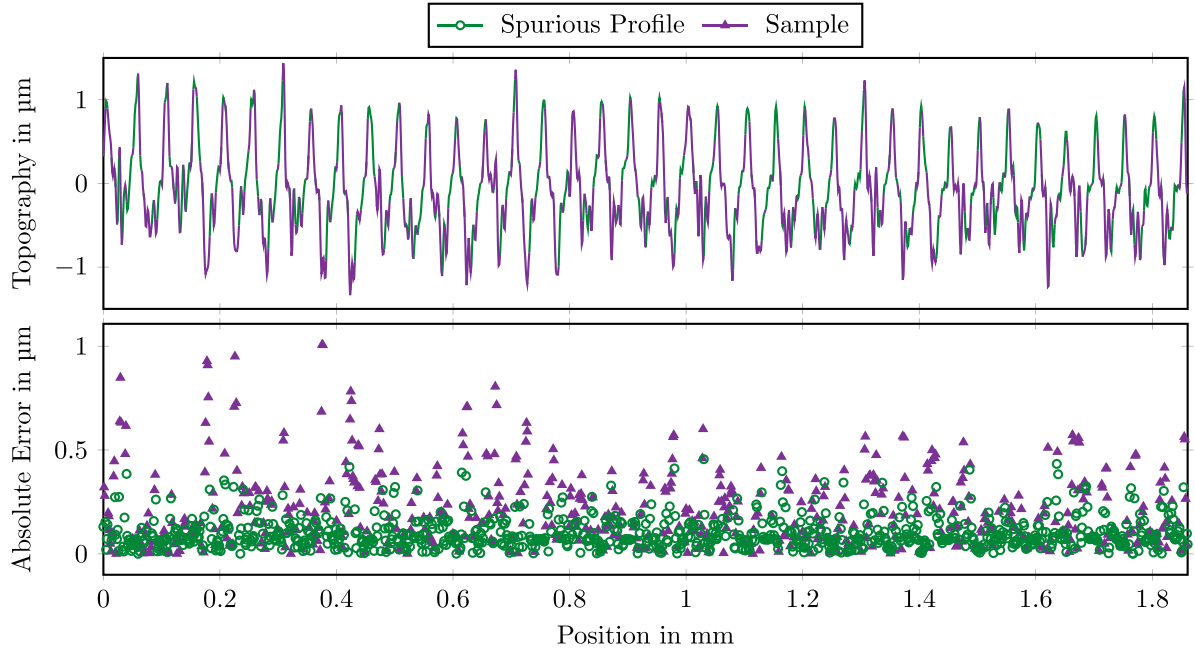

**Figure 8.** The top row shows the spurious measurement of a turned profile, reconstructed with a sample from the posterior GP. The bottom row shows the absolute error between the reconstructed profile and the reference profile. Note that the errors are categorized based on their origin: those from the spurious profile and those from the sample.

The model parameters were optimized based on the spurious profile, and the final model was used to sample from the posterior GP for surface data imputation. The reconstructed profile is shown at the top of figure 8. For simplicity, the explicit posterior GP, including its expectation and a confidence interval, is not shown. The reconstructed profile has characteristics similar to the reference profile (see the top row in figure 7). Notably, it replicates the same number of profile features as the reference profile, even though the model is based on a profile with a relatively high number of spurious points.

Despite this, the imputation frequently underestimates or overestimates the reference heights at spurious locations (see bottom row in figure 8). There are several reasons for this erroneous behavior. First, the model primarily captures spatial properties through the autocovariance function, while height characteristics are only implicitly influenced by the model and the dataset. Thus, prior information based solely on the root mean square height is insufficient for accurate reconstruction of the profile heights. Another source of errors emerges from the differences between the spurious profile and the reference profile. The spurious profile is not directly comparable to the reference profile (see figure 8). Although only one post-processing step was modified to introduce more spurious points, the spurious profile deviates from the reference profile. For example, the spurious profile has a higher topography height than the reference profile at the end of the profile (see figure 7). This problem highlights the challenge of making measurements fully comparable, and the errors presented in figure 8 should be interpreted with caution. The deviations of the sample from the reference are larger than those of the spurious profile from the reference. However, if we consider the errors of the sample as a consequence of the errors of the spurious profile, then the deviations of the sample seem smaller.

## 4. Conclusion

We applied stochastic processes for the filling of missing values in surface data–surface data imputation. In this paper, we assumed that a surface can be described by GPs [20]. Under this assumption, we demonstrated that using stochastic processes for surface data imputation can be advantageous as it allows for the incorporation of expressive prior knowledge about the surface.

We applied our approach to a turned profile, where we included information about the profile feature spacing. Even though complete features were missing, our method was able to reconstruct them. In the second use case, we addressed a chirp-structured profile, incorporating knowledge about the increasing wavelengths. In this problem, our model was able to estimate the missing values with changing wavelengths. Finally, we demonstrate the applicability of our method to a real turned profile. However, validation was challenging because the reference surface already differed from the spurious measurements. Despite this, the reconstructed profile resembled the reference profile, indicating the potential of our approach for real-world applications.

The suggested method generally scales poorly with the size of surface measurement data. Therefore, approximations are needed for computational efficiency; nevertheless, significant work has already been done to make GPs more scalable. Another downside is that the less prior knowledge is available about the surface, the worse our approach becomes. When less information is present, it is debatable whether our approach is generally better or worse than other methods, as it strongly depends on the specific data imputation problem. If we aim to estimate missing values in spurious surface data without any knowledge of the missing values, we state that appropriate solutions for data imputation may not be feasible.

For future work, a comparative study can be conducted to evaluate the effectiveness of this approach for different types of spurious points. In addition, our objective is to conduct a comprehensive study comparing our method with other data imputation techniques to assess its strengths and limitations in various scenarios.

Potential improvements include using heteroscedastic Gaussian noise models that maintain the stationarity of the ground structure while allowing the overall model to be non-stationary. This could help mitigate jumps in the transitions between spurious and valid data. Furthermore, integrating a transfer characteristics model could enhance the inference of the true surface structure. This would correct for the distortions introduced by the measurement process.

Furthermore, this method can be applied not only to estimate missing values, but also to detect spurious points, as it makes inferences by incorporating knowledge about the existing surface.

## Data availability statement

The data cannot be made publicly available upon publication because they are needed to support ongoing research. The data that support the findings of this study are available upon reasonable request from the authors.

## Acknowledgments

This research was funded by the Deutsche Forschungsgemeinschaft (DFG, German Research Foundation) under Grant Number 511263698 (TRR 375/1, Project C03).

## ORCID iDs

A Jawaid  https://orcid.org/0000-0001-9951-4164
S Schmidt  https://orcid.org/0000-0001-7527-3423
M Lotz  https://orcid.org/0009-0000-2706-2111
J Seewig  https://orcid.org/0000-0002-1420-1597